# Analysis of the application of the optical method to the measurements of the water vapor content in the atmosphere – Part 1: Basic concepts of the measurement technique

V. D. Galkin[2], F. Immler[1], G. A. Alekseeva[2], F.-H. Berger[1], U. Leiterer[1], T. Naebert[1], I. N. Nikanorova[2], V. V. Novikov[2], V. P. Pakhomov[2], and I. B. Sal'nikov[2]

[1]Deutscher Wetterdienst, Meteorologisches Observatorium Lindenberg – Richard-Aßmann-Observatorium, Lindenberg, Germany

[2]Russian Academy of Sciences, The Central Astronomical Observatory at Pulkovo, Saint-Petersburg, Russia

**Abstract**

We retrieved the total content of the atmospheric water vapor (or Integrated Water Vapor, IWV) from extensive sets of photometric data obtained since 1995 at Lindenberg Meteorological Observatory with star and sun photometers. Different methods of determination of the empirical parameters that are necessary for the retrieval are discussed. The instruments were independently calibrated using laboratory measurements made at Pulkovo Observatory with the VKM-100 multi-pass vacuum cell. The empirical parameters were also calculated by the simulation of the atmospheric absorption by water vapor, using the MODRAN-4 program package for different model atmospheres. The results are compared to those presented in the literature, obtained with different instruments and methods of the retrieval. The reliability of the empirical parameters, used for the power approximation that links the water vapor content with the observed absorption, is analyzed. Currently, the total (from measurements, calibration, and calculations) errors yield the standard uncertainty of about 10% in the total column water vapor. We discuss the possibilities for improving the accuracy of calibration to ~1% as indispensable condition in order to make it possible to use data obtained by optical photometry as an independent reference for other methods (GPS, MW-radiometers, lidar, etc).

1. **Introduction.**

Atmospheric water vapor is the most important trace gas in the atmosphere, since it plays the key role in its energy budget, the water cycle, cloud formation, and precipitation, as well as in the greenhouse properties of the atmosphere. Due to its large temporal and spatial variability, its observation still poses great challenges to experimentalists.

The optical method for measurements of the atmospheric water vapor content has already been used for almost a century (Fowle, 1912, 1913, 1915). In order to determine the total column water vapor content (or Integrated Water Vapor, IWV), the absorption caused by the water molecules in the near-IR spectral region is measured by a radiometer, with the use of the Sun or a star as a light source. The inverse problem, i.e. the retrieval of the water vapor content from the measured value of the absorption, requires careful calibration. Although other methods for the observation of the atmospheric water vapor content were also introduced later (such as radiosondes, microwave radiometry, and GPS delay), the optical method has not lost its value. Numerous sun photometers designed for studying aerosol components of the atmosphere contain a channel aligned on water vapor absorption bands, which makes it possible to observe quantitatively the water column in the atmosphere (Michasky et al., 1995; Halthore et al., 1997; Schmid et al., 1996; Ingold et al., 2000; Leiterer and Weller, 1988; Leiterer et al.,1995; Alekseeva et al., 1995; Leiterer et al., 1998; Rollin, 2000; Morys et al, 2001; Campmany et al., 2010; Schneider et al., 2010; Pérez-Ramirez et al., 2008; Pérez-Ramirez, 2011). Since 1995, Lindenberg Meteorological Observatory has routinely been monitoring the aerosol optical thickness at standard wavelengths using sun and star photometers, and the water vapor content has also been retrieved from these measurements (Leiterer et al., 1998, 2001; Alekseeva et al., 2001; Novikov et al., 2010). Analysis of the internal uncertainty of these data is presented in Galkin et al. (2010).

Previously, the water vapor content in the atmosphere was observed at night-time by astronomical methods, in order to reveal and analyze variations in the atmospheric transparency in the regions of telluric water vapor bands during astronomical observations (Galkin and Arkharov, 1980, 1981; Alekseeva et al., 1983). When using a filter centered on a water vapor absorption band, the measured flux is affected by the absorption in the set of lines with different degrees of saturation and values of the parameters describing the absorption in the line. Therefore, the dependence of the absorption by water vapor on the water vapor content should be determined empirically. Generally, to describe the dependence of the optical thickness in the absorption band on the water vapor total column and pressure, the power function is used (Golubitskyi and Moskalenko, 1968; Moskalenko, 1968, 1969). A similar approach was used at Pulkovo Observatory, in the process of compilation of the Pulkovo Spectrophotometric Catalog, to retrieve extraterrestrial brightnesses of stars in the spectral regions affected by telluric contamination (Galkin and Arkharov, 1980, 1981; Alekseeva et al., 1997). The necessary empirical spectral parameters of the power function were obtained with the SF-68 spectrophotometer and the unique Pulkovo multipass vacuum cell VKM-100 within the interval of the water vapor content 0.3 - 5.0 cmppw (cm of pre precipitable water) (Alekseeva et al., 1994). On the basis of these data and individual spectral transmission curves for filters used in star and sun photometers, the empirical parameters can be calculated and subsequently used to determine the atmospheric water vapor content from observations with a particular instrument.

In recent publications, water vapor absorption spectra were calculated on the basis of radiative transfer models (e.g., LOWTRAN or MODTRAN), in order to obtain the empirical parameters of the power function (Michasky et al., 1995; Halthore et al., 1997). Schmid et al. (1996) and Ingold et al. (2000) determined the empirical parameters from the comparison of photometrical data (obtained with a sun photometer) with the measurements of the atmospheric water vapor made with microwave radiometers or radiosondes. These empirical parameters differ noticeably from those calculated within the models.

While the total standard uncertainty of the water vapor content obtained by the optical method is about 10 % (Schmid et al., 1996; Ingold et al., 2000), the error of one photometric measurement itself is only about 0.5%, (Galkin et al., 2010). The loss of accuracy during of procedure of the water vapor retrieval is first of all a problem of the theoretical or experimental way of defining the calibration dependence between the absorption for the given filter and the water vapor content in the line of sight.

Therefore, our goal was to check the reliability of the calibration on the basis of laboratory modeling for the absorption by atmospheric water vapor with the use of the VKM-100 multipass vacuum cell. In this cell, a variation of the absorption by water vapor can be accurately related to the variation of water vapor content along the line of sight which is attained by varying the number of passages of the light through the cell. This makes it possible to study the form of the approximation for the relative calibration dependence on the water vapor content (in relative units of the number of passages), for various values of the pressure and temperature, with the standard uncertainty ~1 %. To this end, numerous measurements were made with the Pulkovo cell to calibrate the Lindenberg's ROBAS-30 sun photometer, star photometer, and high-resolution ASP-12 spectrograph. It is possible to derive the absolute calibration from measurements of the humidity in the cell, which are currently made with polymer sensors of a limited accuracy.

In addition, the calibration (i.e. the determination of the empirical parameters) was made on the basis of the Pulkovo Catalogue (Alekseeva et al., 1997), and also of calculated spectra taken from the MODTRAN-4 database. Further on, we used radiosonde data to calibrate our photometers. These methods are described in Chapter 3. The results obtained from these approaches are presented in Chapter 4 and discussed in Chapter 5. On the basis of these studies, we have developed some ideas to improve the reliability of the photometric method, in order to fully explore the potential of this technique as an independent reference for determination of the atmospheric water vapour content.

## 2. The optical method

## 2.1 The empirical approximation for the absorption in the water vapor spectrum.

Since more than 90 % of measurements made with photometers are carried out when the amount of the water vapor along the line of sight is within the interval 0.5 - 5.0 cmppw, the absorption in this interval should be calculated or obtained experimentally. At a certain moment, the effective pressure and temperature for the atmospheric water vapor deviate from their average values by no more than 5 %. In the optical method, the absorption in the certain interval of wavelengths is averaged by the filter or a slit of the spectrophotometer over the spectral lines which lie within the given interval. In addition to the absorption in multiple lines within the wavelength interval of the used filter, the observed signal value is also influenced by Rayleigh scattering and aerosol absorption. In the course of observations, all these factors are taken into account routinely; this procedure is described in detail in (Alekseeva et al., 2001; Novikov et al., 2010; Galkin et al., 2010).

According to the statistical model, the absorption in multiple spectral lines is given by the expression (Goody, 1964)

$$A = 1 - T = 1 - \exp\{-(1/\Delta\lambda) \Sigma W_i\}, \quad (1)$$

where $A$ is the absorption in multiple spectral lines, $T$ the transmission, $W_i$ the equivalent width of the $i$-th line, $\Delta\lambda$ the wavelength interval. The expression (Eq. 1) makes it possible to calculate the absorption in multiple lines depending on the pressure, the temperature, and the amount of water vapor, provided the spectroscopic parameters of the individual lines are known. The expression (Eq. 1) does not yield the analytical dependence of the absorption on the number of absorbing water vapor molecules $W$, the pressure $P$, and the temperature of the vapor. More promising is the empirical approach to the determination of this dependence (at least from the physical parameters $W$ and $P$) based on the approximation of the variations of the optical depth $\tau$ as a function of the water vapor content $W$ and the pressure $P$ by a power law:

$$T = \exp(-\tau) = \exp\{-\beta \cdot W^\mu \cdot P^n\}, \quad (2)$$

were $\beta, \mu, n$ are empirical parameters. Note that (Eq. 2) contains separate dependences on $W$ and $P$, with different power indices. The temperature dependence of the optical depth can be included in the parameter $\beta$. However, the influence of the temperature on the transmission does not exceed 1-2 % for the temperature interval in the atmosphere, and can therefore be neglected in the first approximation. In operations with star and sun photometers, star magnitudes $m$ are commonly used, defined by the following relation: $m = -2.5 \lg(I)$, where $I$ is the intensity of the optical star radiation (in $W \cdot m^{-2}$). Therefore, the absorption by water vapor in terms of the star magnitudes is, according to (Eq. 2):

$$[m-m_o](W) = -2.5 \lg(T) = 2.5 \lg(e) \cdot \beta \cdot W^\mu \cdot P^n = c \cdot W^\mu \qquad (3),$$

where $[m-m_o](W)$ is the absorption in a water vapor band ($m$ and $m_o$ are the star magnitudes with and without the absorption, respectively), $c$ and $\mu$ are the empirical parameters that describe the absorption at a given pressure. In particular, $c$ is the absorption by water vapor in star magnitudes for 1 cmppw, and $\mu$ - the dimensionless parameter describing the variation of the absorption with the water vapor concentration. The parameter $c$ is constant for a given pressure. In real observations, in the first approximation it corresponds to the average effective pressure of the water vapor in the atmosphere, $P_{eff}$ (for Lindenberg, $P_{eff} = 0.845$ atm = 856 hPa). Estimates show that $P_{eff}$ deviates from its average value by less than ±70 hPa, which corresponds to the expected maximum variation of $c \pm 4$-5 %. Therefore, the dependence of the parameter $c$ on $P_{eff}$ should be taken into account only for a small number of abnormal cases.

Schmid et al. (1996) and Ingold et al. (2000) used similar approximations for the dependence of the transmission on the amount of the water vapor and calculated the empirical parameters using MODTRAN. In these cases, the empirical parameters $a$ and $b$ are related to our parameters $c$ and $\mu$ as follows:

$$a = c / 2.5 \lg(e) = 0.921c; \quad b = \mu \qquad (4)$$

In the literature, other approximations of the absorption in multiple spectral lines were also discussed, in the form of a combination of trigonometric functions or polynomials of different degrees. Some of them are reviewed in (Golubitskyi and Moskalenko, 1968; Moskalenko, 1968, 1969). However, in practice the approximation by power function (2) is preferred. This approximation successfully represents the dependence of absorption on the concentration of the absorbent, pressure and temperature as a product of functions of these parameters. However, the disadvantage of power approximation is that it insufficiently accurately represents the calculated or experimental dependence of absorption on the amount of the absorbent. Therefore, the values of the empirical parameters depend, in particular, on the interval of the contents of the water vapor, within which the approximation is carried out. Thus, further studies are necessary to obtain a more accurate form of the approximation (2); for example, different parameters may be used in the expression (2) for different intervals of water vapor contents, or another analytical form of the approximation may be searched.

## 2.2 The usage of the multipass vacuum cell for determination of the empirical parameters c and μ

The water vapor absorption was studied with the use of the VKM-100 multipass vacuum cell, in which the system of mirrors was placed according to White's scheme (Galkin et al., 2004; White, 1942).

Fig.1a presents the general optical schematic diagram of the cell.

The spherical mirrors A, B, and C with the radius of curvature 96.5 m are mounted so that the mirrors A and B form a consecutive set of images of the entrance slit on the mirror C. The mirror C reflects the mirror A onto the mirror B, and vice versa. The input objective O1, located in the plane of the entrance slit E, reflects the light source (restricted by the diaphragm S) onto the mirror A. The diaphragm S restricts the size of the light beam to the solid angle of the mirror A, thereby eliminating superfluous light scattering in the cell. The number of light passages varies due to variation of the relative position of the optical axes of the mirrors A and B and, hence, to variation of the number of images on the mirror C. We can see in Fig.1a that the mirrors A and B should be adjusted so that in the upper row of images formed on the mirror C, an odd number of images is formed; given that, the last (even) image will be placed on the exit slit. In contrast to White's scheme, instead of the exit slit, the mirror D is introduced, which reflects the mirror B onto the output objective O2. Thereby, the system of mirrors A, B, and C, makes it possible to obtain multiple passages of light, starting with the minimum number of passages equal to 4, and then increasing it by an integer factor. Thus, the images of the entrance slit appear in the exit window of the cell (behind the objective O2) after the number of passages equal to 5 (4+1), 9 (8+1), 13 (12+1), 17 (16+1) etc. The maximum number of passages is restricted by the number of images of the entrance slit which can be placed along the mirror C (for the VKM-100 cell, this number reaches a hundred images, which corresponds to the length of the path of 40 km). However, in practice, the maximum number of passages is substantially lower due to light losses on reflection, which vary as $r^N$, where $r$ is the reflection index, and $N$ the number of reflections. For the path length of 4100 m, the signal decreases by 6 star magnitudes (by the factor of 250), which corresponds to the reflection index of mirrors ~ 89 % (aluminum covering). Another reason limiting the maximum distance that light can pass in the cell is the diffusion of the entrance slit image with the increase in the number of reflections. This is due to insufficient quality of the surfaces of the mirrors caused by difficulties with the testing of the curvature radius for mirrors with such small curvature. Improving the quality of mirror surfaces and using silver covering (with the reflection index 95-96 %), one may substantially increase the maximum number of light passages and the corresponding interval of contents of water vapor along the line of sight.

The length of the cell is 97.5 m; the minimum path length used for our measurements was 500 m. The measurements were also made with the path length 900, 1300, 1700, 2100, 2500, 2900, 3300, 3700, and 4100 m. Figure 1b presents the general structure of the experiment.

The amount of water vapor along the line of sight depends on the path length and the absolute humidity in the cell. The latter was measured by four polymer sensors connected with the control unit; the data obtained from the sensors were periodically logged in and averaged. A detailed study of these sensors for various values of relative humidity, temperature, and pressure was carried out at Lindenberg Meteorological Observatory. Our sensors were calibrated to the standard humidity of saturated vapor above various salt solutions and also to the data obtained with TOROS reference devices used for measurements of humidity at the frost point and by Vaisala sensors that used the FN technique introduced at Lindenberg Observatory (Leiterer et al., 1997). A comparison between our sensors and reference instruments in a climate chamber was carried out in Lindenberg by Galkin et al. (2006) and showed that the uncertainty of the measurements of humidity in our cell was only 5-10 %.

For several years, the calibration of star and sun photometers with the VKM-100 cell was made in accordance with the scheme in Fig.1b. The water vapor content, as a rule, was determined by the polymer sensors. Some of the calibrations of the photometers with the VKM-100 cell were accompanied by measurements made with the high-resolution ASP-12 spectrograph. The equivalent width of the water vapor absorption line at 694.3803 nm was determined (see Fig.1c). The measurement of the equivalent width of this line makes it possible to determine the water vapor content in the cell at various pressure. These measurements made it possible to determine the water vapor content under conditions of low relative humidity (<30-40%), when the measurements with polymeric sensors were unreliable (Galkin et al., 2006).

Later on, ASP-12 and the sun photometer ROBAS-30, calibrated under the same conditions in the cell, were used for simultaneous determinations of atmospheric water vapor content made from observations of the Sun. The purpose of these observations was to compare the experimental data obtained by two optical methods for identical light paths in the atmosphere. The first instrument used an isolated absorption line, with the intensity independent of the temperature, while the second analyzed a set of lines with different intensities and temperature dependence.

The comparison between the two techniques of observations depends not only on specific features of the accepted methods, but also on imperfections of the used photometers. This is the reason why, to discriminate the sources of errors, we carried out our observations simultaneously. The results of the comparison of the photometers will be considered in detail in a separate study.

Figure 2 presents the relative spectral transmission curves of the interference filters used in Lindenberg star and sun photometers, and the spectral distributions for the parameters $c$ and $\mu$ in the region of 935 nm water vapor absorption band (Alekseeva et al., 1994). Fig.2 displays a high degree of variability of the spectral parameters $c$ and $\mu$ within the broad wavelengths interval of the sun filters. Using the data presented in Fig.2, it is possible to calculate the parameters $c$ and $\mu$ for any given filter.

Measurements of the intensity of light that passed through the cell were carried out with star and sun photometers (Fig.1b). Figure 3 presents an example of such measurements made with the BAS-30 sun photometer, with the path length of 2500 m through the air with $P = 0.9$ atm and through an evacuated cell ($P = 0.001$ atm). The ratio of intensities of the spectra observed with the filled and empty cell yields the water vapor transmission for the given path length. The transmission obtained for another path length indicates the variation of the transmission with the increase of the water vapor content along the line of sight (Fig.4). In Fig.4, the measured transmission (in star magnitudes) is presented as a function of the length of the light path in the cell (in the units of the minimum path, 500 m). This illustrates the variation of the transmission in on-sky measurements with the increase of the zenith distance of the observed object. The approximation of the data in Fig.4 by a power function (3) yields the values for the parameters $c$ and $\mu$ with the standard deviations $\sigma_c = 0.004$ and $\sigma_\mu = 0.014$.

The procedure of measurements of the parameters does not last longer than half an hour, which provides an opportunity to study the dependence of the empirical parameters on the conditions in the cell (the temperature, pressure, water vapor content, and path length). The error of determination of the parameter $c$ is caused primarily by the error of the sensors used for the measurements of the absolute humidity. On the other hand, the given technique makes it possible to carry out further experiments in order to increase our level of knowledge about the absorption of water vapor and various forms of approximations for the absorption as a function of the water vapor content.

**3.     The results of determination of the empirical parameters with different methods.**

The study of absorption by water vapor under various physical conditions makes it possible to consider separately the dependence of absorption on the amount of absorbing substance, pressure, and temperature. The dependence of absorption on the amount of the water vapor along the line of sight for constant pressure and temperature is established easily (varying the number of passages of light in the cell); however, the variation of pressure or temperature for constant humidity presents more serious experimental problems.

The measurements of the parameters $c$ and $\mu$ were carried out with the star photometer and sun photometer BAS-30 for pressures ranging from 0.1 to 1 atm with the step of 0.1 atm. Table 1 presents the results obtained for one of the filters (948.0 nm) with the star photometer for various pressure values.

It follows from Table 1 that the parameter $\mu$ only weakly depends on pressure. This justifies the assumption of separate dependences of the absorption on pressure and concentration. Parameter $c$ corresponds to the amount of water vapor in the minimum path length 500 m. To recalculate the parameter $c$ for 1 cmppw, we used the readings of the humidity sensors, which resulted in the increase of the determination error for the parameter $c$. Table 1 also contains standard deviations for the determined parameters. They are specified only by the accuracy of the photometric measurements, the stability of the source of radiation, and the adjustment of the optical scheme.

Table 2 presents the results obtained for other water vapor- centered filters of the star photometer and the BAS-20 and BAS-30 sun photometers, and laboratory data for the SF-68 spectrophotometer (taken from Alekseeva et al., 1994). Columns 1 and 2 contain the central wavelengths of the water filters and their widths; columns 3 and 4 - parameters $c$ and $\mu$ obtained from direct measurements in the cell. All values of the parameter $c$ were recalculated for 1 cmppw and the pressure 0.845 atm, corresponding to the effective pressure of the water vapor in the atmosphere at the sea level.

The columns 5 and 6 present $c$ and $\mu$ calculated from the spectral transmission functions (according to Alekseeva et al., 1994) and from the transmission curve of the water vapor-centered filter of the photometers. The columns 7 and 8 contain $c$ and $\mu$ calculated with MODTRAN-4. The last columns (9 and 10) present $c$ and $\mu$ obtained from on-sky observations with sun and star photometers calibrated by radiosonde data. In practice, parameters for Lindenberg sun photometers were determined as follows. For a given photometer and for a particular water vapor-centered filter, the empirical parameters $c$ and $\mu$ were derived from the laboratory spectra obtained at Pulkovo with the VKM-100 vacuum multipass cell and the SF-68 spectrophotometer within the range of water vapor contents 0.5 - 5.0 cmppw along the line of sight (Alekseeva et al., 1994). Then the parameter $\mu$ (column 6) was used to derive the extraterrestrial magnitude $m_0$ of the Sun in the water vapor band and the parameter $c$, from radiosonde and observational photometrical data, for the time interval when the calibration of the photometers did not vary.

Figure 5 presents the example for the calibration curve for the ROBAS-20 sun photometer, the 945.51 nm filter, and the observational period from April, 2002 to August, 2003. The observational $(m_{obs.} - (\alpha_{Ray} + \alpha_{aer.}) \cdot M)$ are plotted as a function of radiosonde data $[W_{RS80} \cdot M]^{\mu}$. Here, $m_{obs}$ is the measured signal from the Sun (in star magnitudes), $\alpha_{Ray}$, $\alpha_{aer}$ - Rayleigh and aerosol components of atmospheric extinction (in star magnitudes), $W_{RS80}$ – the water vapor content derived from radiosonde data (cmppw), $M$ – the air mass. Figure 5 demonstrates that the obtained 7154 individual measurements are closely matched with a linear dependence. For the other observational periods, the sun photometers ROBAS-20 and ROBAS-30 were calibrated in a similar way. For the star photometer, the procedure of determination of the parameters $c$ and $\mu$ was slightly different, however, the basic principle of the selection of the parameter $\mu$ on the basis of laboratory data and the recalibration of the parameter $c$ according to radiosonde data was maintained.

The total volume of observational data obtained with the star and sun photometers from the year 1995 to 2008 was processed with the parameters determined as it was described above and presented in Fig.6. Ground-based GPS-receivers deliver continuously data for climate and NWP (Numerical Weather Prediction) applications. In order to ensure the high quality of this products reference data for Integrated Water Vapor (IWV) from independent instruments are required for quality control and accuracy estimation. To meet the general needs for high-quality water vapor information the WCRP/Global Water Vapor Project (GVaP) was initiated, which includes the establishment of reference observation stations. The *Deutscher Wetterdienst* is setting up the *Meteorologisches Observatorium Lindenberg* as validation site. Also, since 2008 this observatory is the first GRUAN (GCOS Reference Upper-Air Network) network station and also hosts the GRUAN lead centre (Seidel et al, 2009; Immler et al., 2010). Lindenberg performs continuous monitoring and validation of IWV using GPS, radiosondes, microwave profiler and 2-channel radiometers since more than 15 years (see for example Güldner, J., 2001). The results of the determination of the water vapor column contents (IWV) made by the optical method and their intercomparison with those obtained with the use of other techniques will be discussed in more detail in a separate publication (*Analysis of the application of the optical method to the measurements of the water vapor content in the atmosphere – Part 2: Intercomparison with data obtained by other devices and techniques*).

**4.    Discussion.**

Table 3 presents the values for empirical parameters for some sun photometers taken from literature. Since the wavelength for the measurement of the atmospheric water vapor content was established by WMO, and most photometers carry out measurements in this wavelength, the parameters obtained in different studies can be compared directly. The parameter $b = \mu$ essentially does not depend on the width and the shape of the transmission curve of the filter; therefore, provided the intervals of water vapor contents are close to each other, the direct comparison of different values for the parameter is possible. The parameter $a = 0.921 \cdot c$ depends on the half-width and shape of the filter transmission curve, and also on the set of data used for the calibration. The height of the point of observations above the sea level also affects the value of parameter $a$.

Figures 7 and 8 present the comparison between our determinations for the parameters $c$ and $\mu$ (Alekseeva et al., 1994 and Table 2) and the data taken from other studies (Table 3). Taking into account that the parameters may be dependent on the used spectral resolution, the data in the Figures are presented as a function of the halfwidth of the transmission curve of the filter. Tables 2 and 3 contain the values of the parameter obtained using different techniques with the same photometers, with photometers of different type, and even with different light sources (the Sun and stars). We present this comparison to demonstrate the consistency of these various types of data and, on the other hand, to find some trends in the variation of the parameters as a function of the half-width of the filter $\Delta\lambda$ (nm). From the total volume of the data (Fig. 7), for the parameter $\mu$ we obtain:

$$\mu = 0.0007 \cdot \Delta\lambda(\text{nm}) + 0.5964 \qquad (5)$$
$$\sigma_\mu = 0.013 \text{ for } \Delta\lambda = 5 \text{ nm}; \; \sigma_\mu = 0.026 \text{ for } \Delta\lambda = 20 \text{ nm}$$

For the parameter c (Fig.8):

$$c = -0.0037 \cdot \Delta\lambda(\text{nm}) + 0.6716 \qquad (6)$$
$$\sigma_c = 0.023 \text{ for } \Delta\lambda=5 \text{ nm}; \; \sigma_c = 0.044 \text{ for } \Delta\lambda=20 \text{ nm}$$

Figures 7 and 8 indicate that, in spite of the differences in the techniques used for determination of the parameters, their consistency is fairly satisfactory. The determination errors for the water vapor contents obtained with the calculated parameters correspond to the standard deviations of water vapor contents measured in real observations, both in the present study and in the studies (Michasky et al., 1995; Halthore et al., 1997; Schmid et al., 1996; Ingold et al., 2000). The error of photometric measurements with a star photometer $0.^m005$ and with a sun photometer $0.^m001$ corresponds to potential possibility to reach uncertainty of measuring of water vapor content up to ~ 1 %. Really, we have a standard uncertainty of the water vapor content of 5-10 % only, both at sun, and at star observations. In our opinion the principal cause of accuracy losses in both cases same is use for definition of extraterrestrial star magnitudes and then for water vapor content definition of (Eq. 3): $[m-m_0] = c \cdot W^\mu$. For example, on Fig. 5 straight line corresponds to the parameters $c$ and $\mu$ obtained in laboratory for range of water vapor contents 0.5-5 cmppw. Points represent data of real measurings with the sun photometer, obtained in Lindenberg during 1.5 years (7154 values). One can see from this figure, that the postulated function well enough features observed data for the basic range of water vapor contents (1 – 9 cmppw along the line of sight, or 1 – 3 for $W^\mu$), where overwhelming majority of points (5857 values) are allocated, and it corresponds to an error of definition of 5-10 %. In too time it is possible to note some diversions of points from a straight line for small and very major water vapor contents, that testifies to insufficient reliability of the accepted approximation. Therefore, $\mu$ depends on the interval of the water vapor contents, and tends to decrease with an increase of the latter. The parameter $c$ depends on pressure. The height distribution of water vapor in the atmosphere varies within a wide range. The variations in the water vapor distribution affect the effective pressure of water vapor and thereby specify the value of parameter $c$. To a larger extent, the parameter $c$ depends on the interval of the water vapor content for which the parameter was determined. The weak dependence of parameter $c$ on temperature also exists. All these factors should be studied and subsequently included into the processing algorithm, to maintain the accuracy of 0.5 % (already reached in photometric observations) and to decrease the error of calibration of the water vapor content closer to 1 %. In order to ensure the metrological traceability and to keep high initial accuracy during data processing, up to obtaining the final values of column water vapor content, it is necessary to analyze carefully the following factors: stability of instrumental photometric system, errors at definition of extraterrestrial star magnitudes, determination of time-trend of atmospheric extinction during observations (especially of the aerosol absorption in the boards of water vapor band), etc. Partially (for old observations) it is made by us in paper (Galkin et al., 2010). We plan to return to this problem in a separate publication (*Analysis of the application of the optical method to the measurements of the water vapor content in the atmosphere – Part 2: Intercomparison with data obtained by other devices and techniques*) devoted to the analysis of the data, obtained in Lindenberg by various devices and methods in 1995-2007. The recommendations in detail for optimization of the solar and stellar observations and data processing algorithms, in order to minimize the final errors in integrated atmospheric water vapor contents (IWP), will be made as Appendixes in this future article too.

A detailed examination of humidity observed in the cell with the use of calibrated sensors showed the impossibility to determine the *integrated* values of the water vapor content in the cell with the accuracy higher than 5 %, using the sensors (Galkin et al., 2006). It is due to both the insufficient accuracy of the sensor readings and the inhomogeneities of water vapor content along the length of the cell, caused by the temperature gradient and local peculiarities. Further on, we plan to use the new thermohygrometer with 4 calibrated polymeric sensors for additional testing the homogeneity of water vapor content along the length of the cell. The data of this testing will make it possible to recalibrate the humidity scale obtained on the basis of Pulkovo spectroscopy by comparison with the standard Lindenberg humidity scale, used for calibration of radiosondes.

In order to calibrate photometric measurements and determine the zero-point of the scale for the empirical parameter $c$, the total (integrated) water vapor content along the total optical way in the cell in absolute units (cmppw) should be known. To this end, we suggest using the ASP-12 vacuum high-resolution spectrograph, with which it is possible to derive the water vapor content from absorption in a separate narrow water vapor line (primarily, 694.3803 nm) with its known half-width and intensity.

The absorption in an isolated spectral line is strictly related to the parameters of the line: its intensity, half-width, and the line shape, and to the physical conditions under which the line is formed (the concentration of the absorbing substance, the pressure and temperature of the absorption mixture). If the physical conditions, under which the measurements are obtained, are known, the line parameters can be determined. And on the contrary, if the line parameters are known, the measurements of absorption in the isolated spectroscopic line can be used for determination of the physical conditions under which the line is formed. The growth curve method (the dependence of absorption on the line parameters and on the physical conditions under which the line is originated) is widely used in astrophysical studies of physical conditions in atmospheres of the Earth, planets, and stars.

In the 70-ies of the last century, the water vapor line with the wavelength 694.38 nm (the parameters of which were repeatedly determined at that time) was commonly used for determination of the water vapor content in the Earth's atmosphere. The typical measurement uncertainty for the intensity and half-width of the line was ~ 10 %.

In the last decade, extensive studies of absorption in water vapor bands at optical and IR-wavelengths have been made (see references in HITRAN 2008 database; Rothman et al., 2009). According to the database, currently there exist a number of lines for which the line parameters have been determined with the uncertainty of 1 %.

The method of determination of the water vapor content from the absorption in a separate narrow line was successfully used at Pulkovo for astrophysical purposes when the Pulkovo spectrophotometric star catalog was being composed (Alekseeva et al., 1997, 1994). However, in that case it was not necessary to know the real absolute water vapor contents in the cell with a very high accuracy. Only the relative homogeneity of scale for the empirical parameter $c$ was really important. As a result, when we tried to apply our previous laboratory spectral tables for parameters $c$ and $\mu$ (Alekseeva et al., 1994) to geophysical instruments (Lindenberg's star and sun photometers), it appeared that our scale of water vapor contents differed systematically from the radiosonde scale (possibly, due to the incorrect zero-point of scale for the parameter $c$). Therefore, we had to correct our water vapor contents scale, recalculating the parameters $c$ with the use of a large volume of interpolated radiosonde data for every year of photometric observations (as it was described above).

In order to transform our optical method of star and sun photometry into a independent reference method for determination of the atmospheric water vapor contents, it is necessary to repeat at Pulkovo the series of spectral measurements for the determination of parameters $c$ and $\mu$ with the VKM-100 – ASP12 laboratory complex, with a substantially higher accuracy. To this end, we are planning to introduce to this complex the new AvaSpec-3648TEC-USB2 laboratory fiber spectrometer.

In order to achieve the desired accuracy ~ 1 %, the optical connection between the exit window of the cell and the entrance window of ASP-12 should be provided with the use of a fiber optic cable (with the length ~25 m). We are also planning to use an improved detection system in ASP-12, with the signal-to-noise ratio of the order of $10^3$.

For more accurate determination of the relative humidity in the cell, a new mirror system should be mounted. Currently, with the reflection factor of the aluminum-coated mirrors less than 89%, on the 4100 m length, the signal is deteriorated for about 6 star magnitudes (250 times). The optical quality of the mirrors is also insufficient. With a new mirror system (with silver coating), the number of light passes could be increased to extend the interval of the measured water vapor content.

The extension of the interval of the measurements, both with the reference to line intensity and to the wavelength interval, will make it possible to involve more lines with various parameters in the measurements of the water vapor content in the cell.

## 5. Conclusion.

Since 1995, at Lindenberg Meteorological Observatory (currently, Richard-Aßmann-Observatory) sun and star photometers have been in operation; using these photometers, we have measured the aerosol optical thickness and atmospheric water vapor content. As a result, a unique database has been formed. To retrieve the water vapor content in the atmosphere from our measurements, we have developed an algorithm based on laboratory data obtained at Pulkovo Observatory with the VKM-100 multipass vacuum cell. Here, we present the empirical parameters that characterize the absorption by water vapor; with these parameters, the amount of water vapor can be retrieved from observations made with different photometers. The parameters obtained by different techniques with these photometers have been compared with data from the literature; it has been shown that the dispersion of the parameter values is consistent with the standard deviation of measurements of water vapor of approximately 10%.

The most efficient way to improve the accuracy is direct calibration of the instrument in the vacuum cell. The advantage of this method is that it does not depend on other methods of determination of the amount of the atmospheric water vapor. The independent calibration would make it possible for the photometric method to become a reference for other instruments. Measurements made with photometers can thus fill the gap in the current practice of atmospheric observations. However, the accuracy of the independent calibration is currently limited by that of the humidity sensors in the vacuum cell, which can be improved, and by the uncertainty in the absolute zero point of our previously tabulated data (Alekseeva et al., 1994). If the absolute humidity in the cell were known with the standard uncertainty of 1% or less, we would be able to provide the calibration for the integrated water vapor content with the standard uncertainty of about 1 % in absolute units (cmppw) too. It should be emphasized that under identical conditions the calibration both for sun and star photometers can be obtained with identical accuracy.

The authors assume that in order to reach the accuracy ~1% in the calibration of the column water vapor content (Integrated Water Vapor, IWV), it is necessary to:
1. obtain more accurate experimental data for the absorption by water vapor within a broad interval of humidity using the modernized VKM-100 – ASP-12 measuring complex at Pulkovo;
2. study the influence of pressure and temperature on the absorption by water vapor and take it into account in the construction of the calibration curve;
3. if necessary, select a more reliable approximation for the experimental calibration curve.

We strongly believe that with the use of our integrated approach, i.e. the combination of laboratory modeling of the absorption, numerical modeling of the atmospheric absorption, and detailed analysis of measurements made with different type photometers, the needed accuracy may be attained.

**Acknowledgments.**

The authors appreciate the German Research Society (DFG, Deutsche Forschungsgemeinschaft) and the Russian Foundation for Basic Research (RFBR) for the long-term support of our works by grants (DFG-projects: 436 RUS 11317612 (R), 436 RUS 113/632/0-1(R), and 436 RUS 113/632/0-2(R); RFBR-project 01-05-04000 NNIO_a).

**Appendix A   Glossary**

**Measurand** - Quantity intended to be measured.

**Uncertainty** - Property of a measurement, characterizing the dispersion of a set or distribution of quantity values for the measurand, obtained by available information. Where possible, this should be derived from an experimental evaluation but can also be an estimate based on other information.

**Standard uncertainty** - Measurement uncertainty expressed as a standard deviation.

**Accuracy** - Closeness of agreement between the result of a measurement or calculation and a true value of the measurand.

**Metrological Traceability** - Property of a measurement result whereby the result can be related to a reference through a documented unbroken chain of calibrations each contributing to the measurement uncertainty.

**PW (Precipitable Water) — (or Precipitable Water Vapor, PWV)** - The total atmospheric water vapor contained in a vertical column of unit cross-sectional area extending between any two specified levels, commonly expressed in terms of the height to which that water substance would stand if completely condensed and collected in a vessel of the same unit cross section.
For this physically the same term the lot of abbreviations is used by different authors using GPS, MW-radiometers, radiosondes (RS), or sun photometry: **PW, PWV, TPW (Total Precipitable Water), IWV (Integrated Water Vapor)** and some more.
In a quantitative sense, most authors use units "cm" or "mm" for PW until today. The National Oceanic and Atmospheric Administration (NOAA) uses for PW the unit "in." (Inches) in its NCDC Radiosonde Database of North America. Only some authors (for example at University of Bern) use SI unit "$kg/m^2$"for PW.

**cmppw** – centimeters pre precipitable water. At *Meteorologisches Observatorium Lindenberg* this abbreviation had been used as unit for column precipitable water **PW** (or integrated water vapor, **IWV**) during long time (Leiterer et al., 1998). In Lindenberg's PW-database (including data obtained by radiosondes, MW-radiometers, GPS, and sun and star photometers) this unit is used also. And at Pulkovo Observatory for laboratory measurements of integrated water vapor content along the line of sight this same unit was used too. Therefore we keep this non-system unit in presented article too.

**Table 1.** The results obtained for one of the filters (948.0 nm) of the star photometer for various pressure.

| P (atm) | c (500 m) | $\sigma_c$ | µ | $\sigma_\mu$ |
|---|---|---|---|---|
| 1.0 | 0.401 | 0.011 | 0.594 | 0.019 |
| 0.9 | 0.370 | 0.010 | 0.561 | 0.018 |
| 0.8 | 0.361 | 0.011 | 0.621 | 0.022 |
| 0.7 | 0.340 | 0.009 | 0.632 | 0.020 |
| 0.6 | 0.325 | 0.007 | 0.581 | 0.015 |
| 0.5 | 0.278 | 0.006 | 0.601 | 0.016 |
| 0.4 | 0.236 | 0.005 | 0.615 | 0.017 |
| 0.3 | 0.212 | 0.005 | 0.603 | 0.019 |
| 0.2 | 0.167 | 0.006 | 0.591 | 0.027 |
| 0.1 | 0.097 | 0.004 | 0.646 | 0.031 |

**Table 2.** Empirical parameters $c$ and $\mu$ obtained with filters centered on the 93 nm water vapor absorption band, with the star photometer, sun photometers BAS-20 and BAS-30, and the SF-68 spectrophotometer (laboratory data Alekseeva et al., 1994).

| 1<br>$\lambda$ (nm) | 2<br>$\Delta\lambda$ (nm) | 3<br>$c$<br>measured | 4<br>$\mu$<br>data | 5<br>$c$<br>from spectrophotometric<br>(Alekseeva et al., 1994) | 6<br>$\mu$<br>data, | 7<br>$c$<br>from MODRAN | 8<br>$\mu$<br>data | 9<br>$c$<br>empirical | 10<br>$\mu$<br>data |
|---|---|---|---|---|---|---|---|---|---|
| spectrophotometer SF-68 | | | | | | | | | |
| 945.0 | 2.5 | | | 0.7235 | 0.5885 | | | | |
| 945.0 | 5.0 | | | 0.7366 | 0.5920 | | | | |
| 945.0 | 10.0 | | | 0.6877 | 0.5940 | | | | |
| 945.0 | 15.0 | | | 0.6497 | 0.6230 | | | | |
| starphotometer | | | | | | | | | |
| 948.0 | 7.0 | 0.615 | 0.6000 | 0.6541 | 0.5946 | 0.5862 | 0.5992 | 0.598 | 0.564 |
| 948.5 | 8.5 | 0.583 | 0.5747 | 0.6390 | 0.5944 | | | | |
| 946.5 | 7.0 | 0.634 | 0.6173 | 0.6529 | 0.5924 | | | | |
| sunphotometer BAS-20 | | | | | | | | | |
| 942.9 | 22.2 | | | 0.6229 | 0.5794 | 0.5724 | 0.5804 | 0.5718 | 0.5794 |
| 945.5 | 22.2 | | | 0.6128 | 0.5775 | | | 0.5478 | 0.5775 |
| sunphotometer BAS-30 | | | | | | | | | |
| 956.8 | 23.5 | 0.384 | 0.584 | 0.4588 | 0.5797 | 0.4431 | 0.5646 | 0.4211 | 0.5797 |

**Table 3.** The empirical parameters used for some sun photometers.

| Author | W(cmppw) | Δλ | a | b | Simulation | Data for calib. |
|---|---|---|---|---|---|---|
| Halthore R.N. et al. (1997) | 0.85 - 23.5 | ~10 nm | 0.616 | 0.594 | MODTRAN-3 | HITRAN-92 |
| | | broad | 0.436 | 0.55 | | |
| Michalsky J.J. et al. (1995) | 0.5 - 6 | 10 nm | 0.344 | 0.578 | MODTRAN-2 | |
| | 0.5 - 25 | | 0.374 | 0.493 | | |
| Schmid B. et al. (1996) | | 5 nm | 0.508 | 0.627 | LOWTRAN-7 | HITRAN-92 |
| | | | 0.546 | 0.621 | MODTRAN-3 | HITRAN-92 |
| | | | 0.549 | 0.629 | FASCOD3P | HITRAN-92 |
| | | | 0.654 | 0.55 | empirical | RS |
| | | | 0.621 (0.022) | 0.591 (0.017) | empirical | MW |
| Ingold T. et al. (2000) | 0.3 - 15 | 5 nm | 0.5681 | 0.5956 | MODTRAN-3.0 | HITRAN-92 |
| | | | 0.5719 | 0.5934 | MODTRAN-3.5 | HITRAN-96 |
| | | | 0.556 | 0.5932 | MODTRAN-3.7 | HITRAN-96 |
| | | | 0.5957 | 0.5984 | LBLRTM 5.10 | HITRAN-96 |
| | 1 - 30 | | 0.6034 (0.0445) | 0.5648 (0.0378) | empirical | MW |

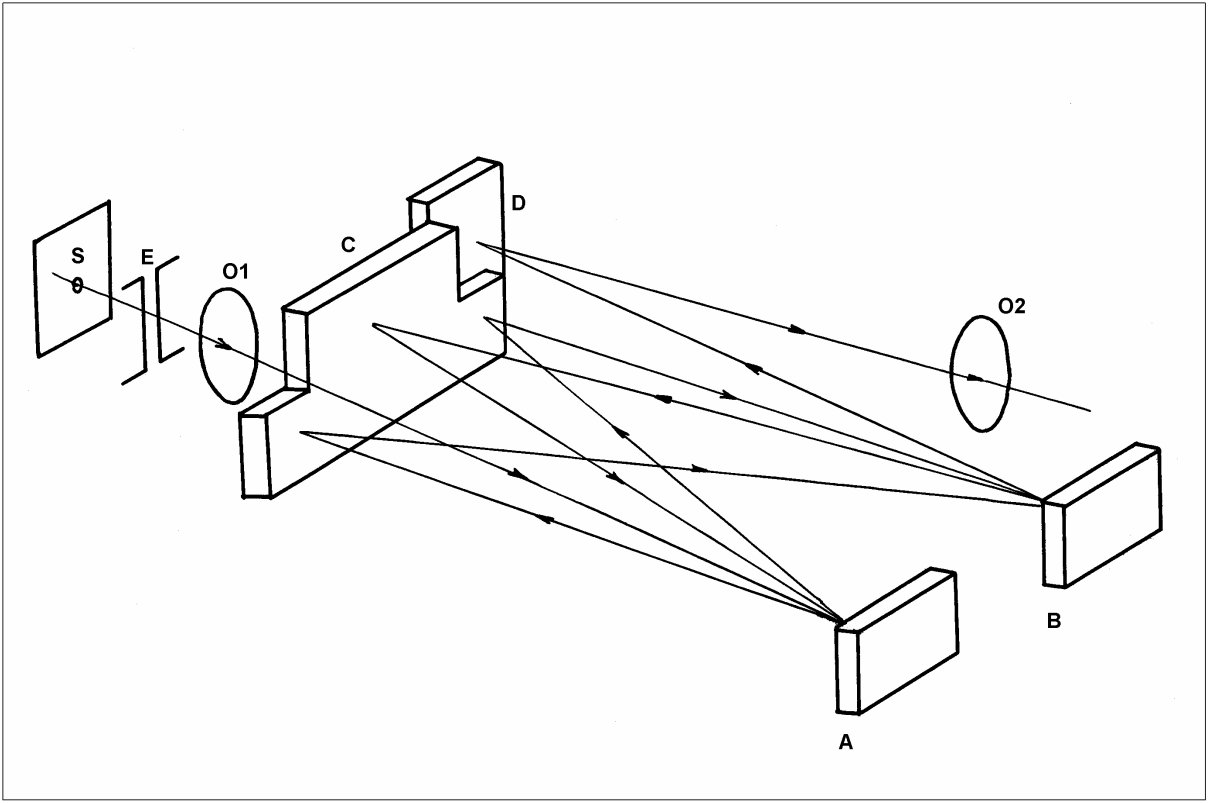

**Fig. 1a**. The general optical schematic diagram of the VKM-100 cell.

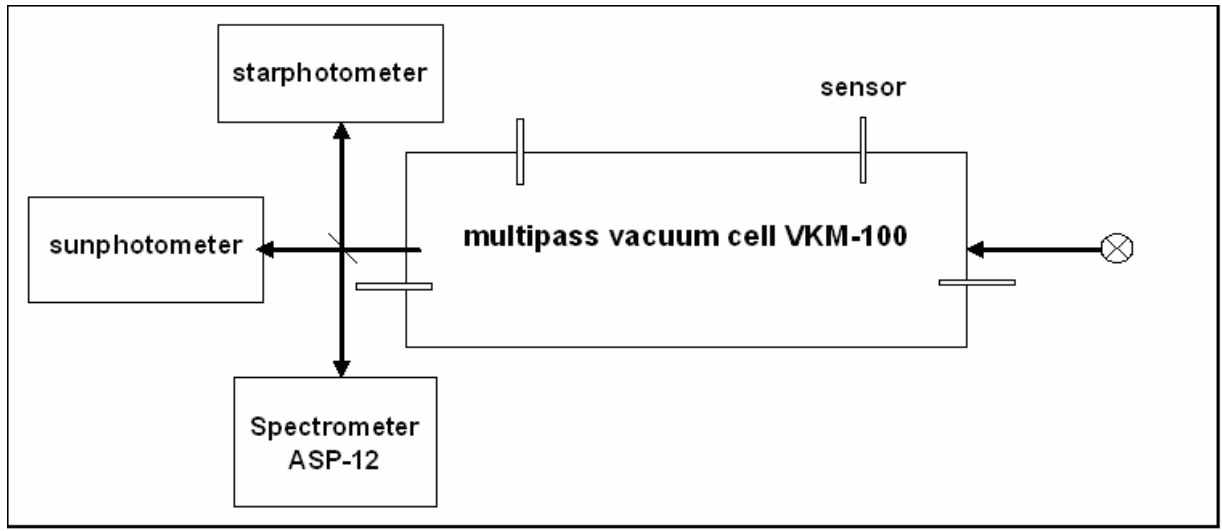

**Fig. 1b**. The general scheme of the set for the calibration of photometers, with the indication of the positions of the humidity sensors.

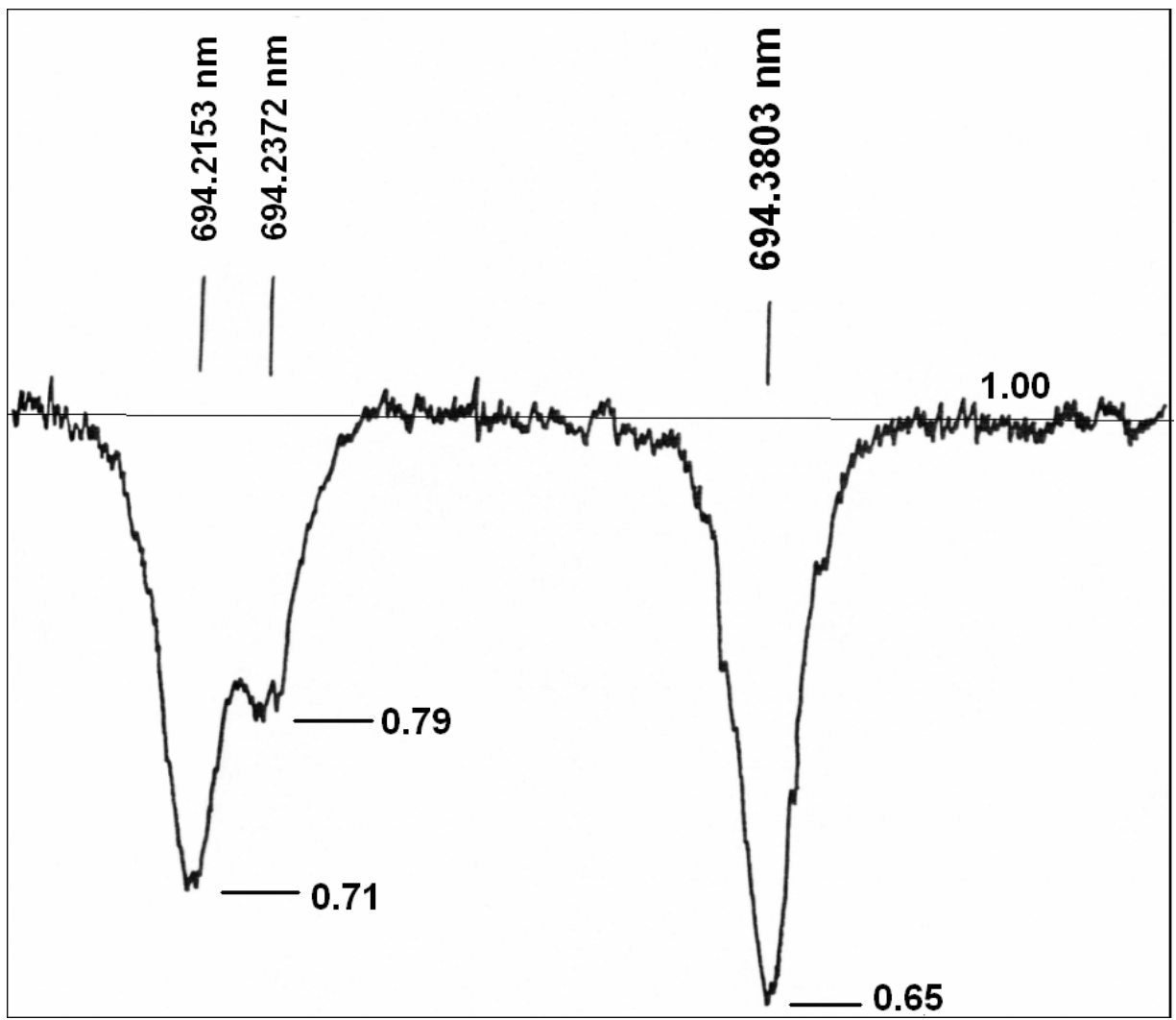

**Fig. 1c.** General view of the absorption spectrum of water vapor in the vicinity of $\lambda =$ 694.3803 nm, obtained with the ASP-12 spectrograph (with the light path length in the VKM-100 cell 1300 m, $W$ = 1.3 cmppw, the slot width 0.01 nm).

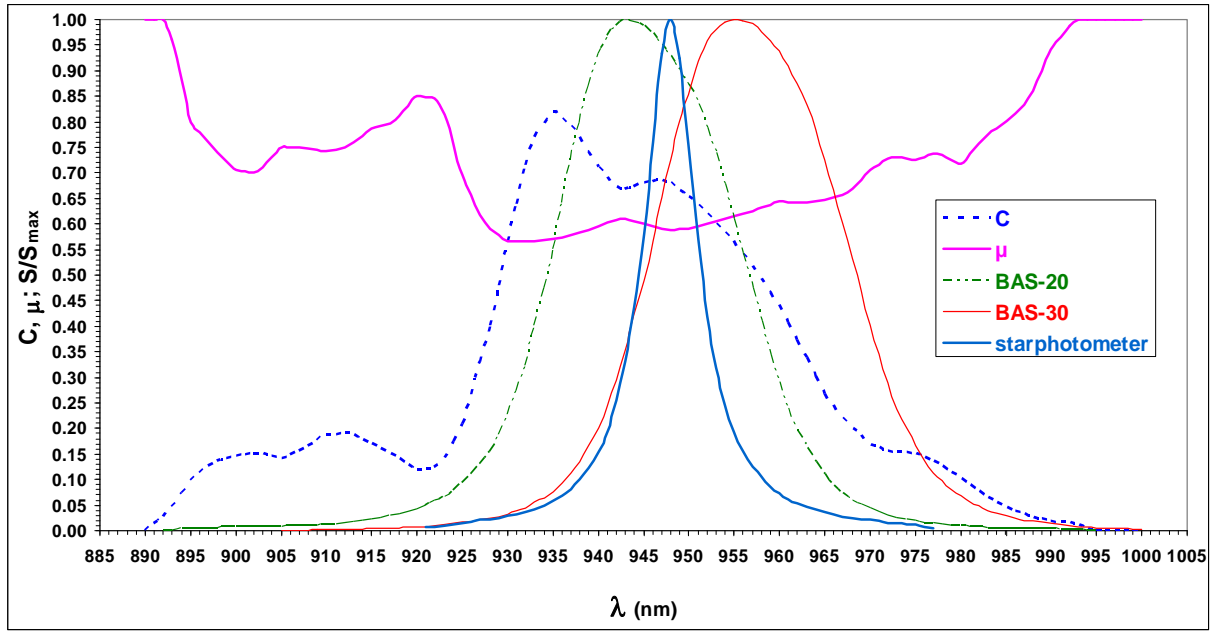

**Fig. 2.** The relative spectral transmission curves for filters of the Lindenberg's star (blue solid curve) and sun (red and green curves) photometers, and the spectral distributions of the parameters $c(\lambda)$ (blue) and $\mu(\lambda)$ (magenta) in the region of 935 nm water vapor absorption band.

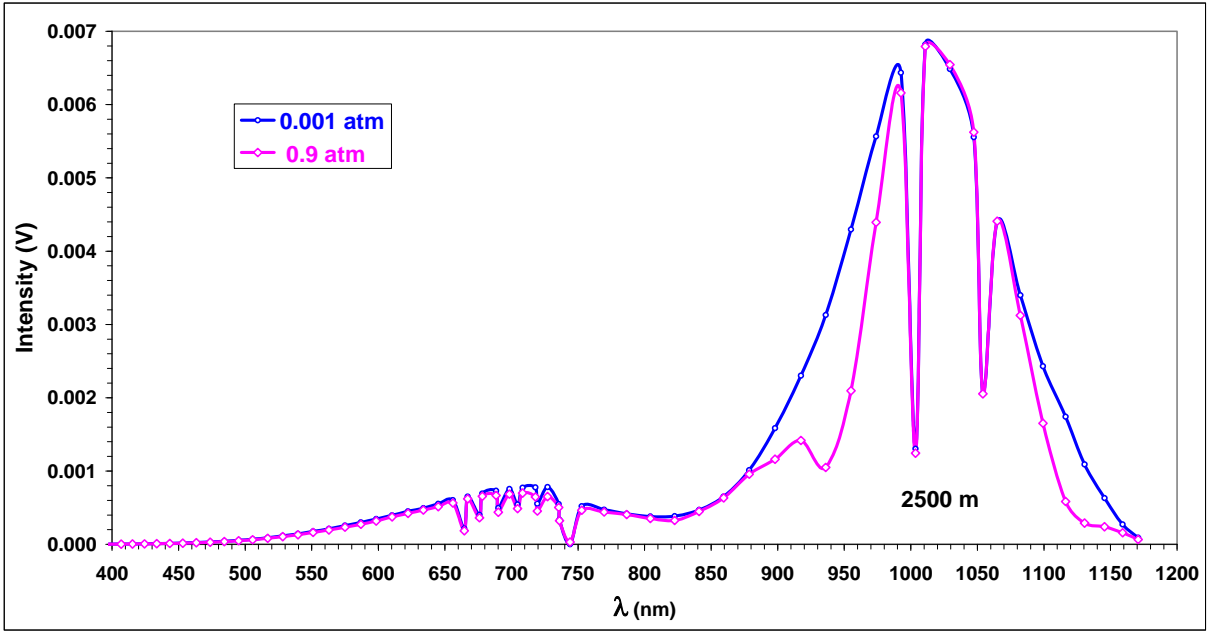

**Fig.3**. The spectra of the light source (lamp) observed with the BAS-30 sun photometer in the evacuated and filled cell.

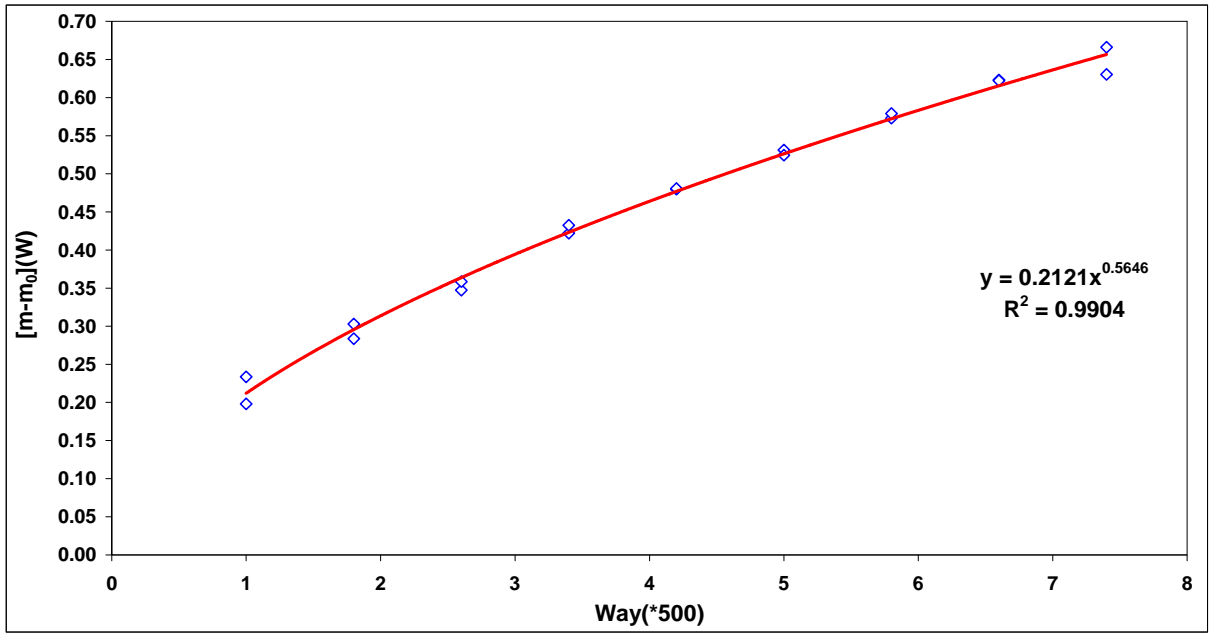

**Fig. 4**. Variation of transmission in the 935 nm water vapor absorption band with the increase of the path length.

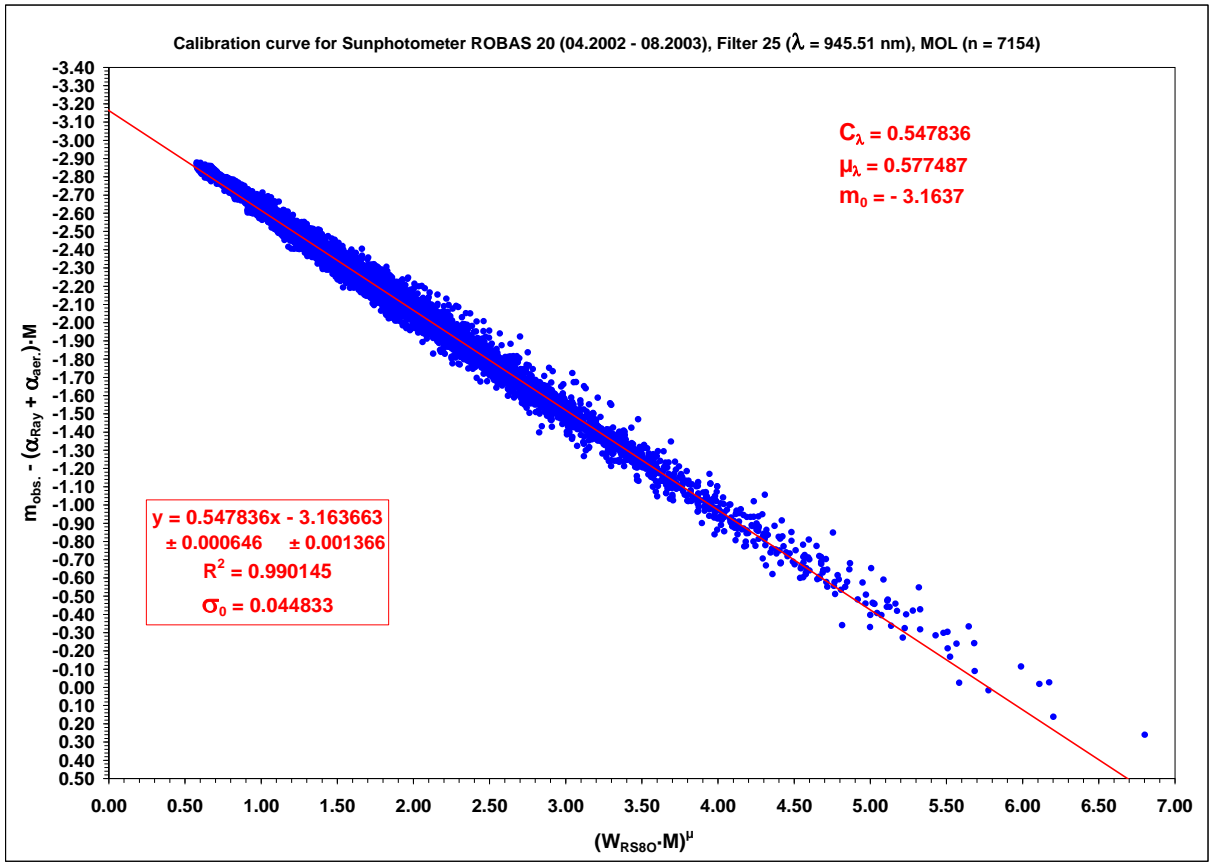

**Fig. 5**. Calibration curve for the ROBAS 20 sunphotometer (04.2002-08.2003), Lindenberg (*n* = 7154).

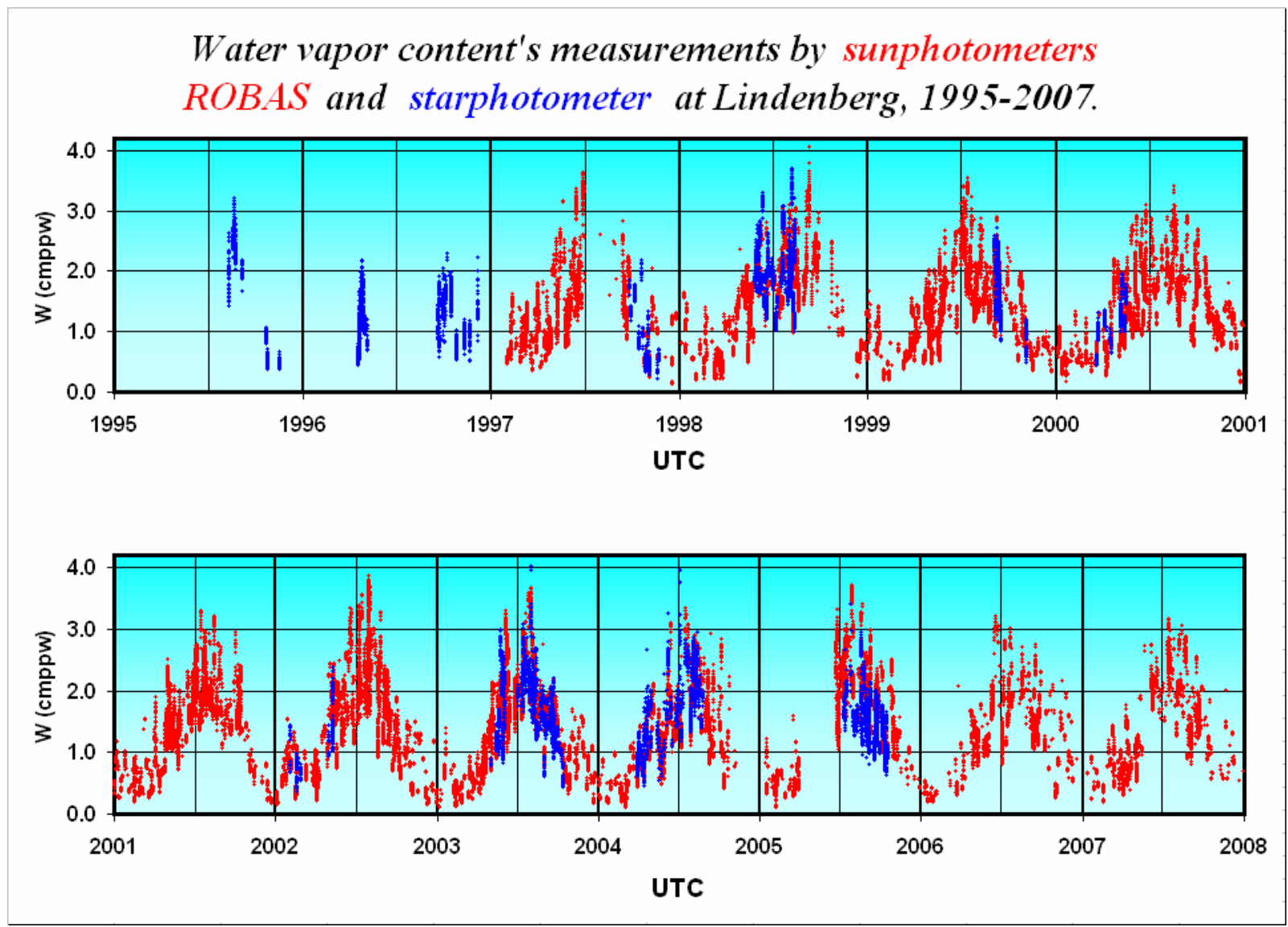

**Fig.6.** The measurements of water vapor content with the ROBAS sun photometers and the star photometer at Lindenberg, 1995-2007.

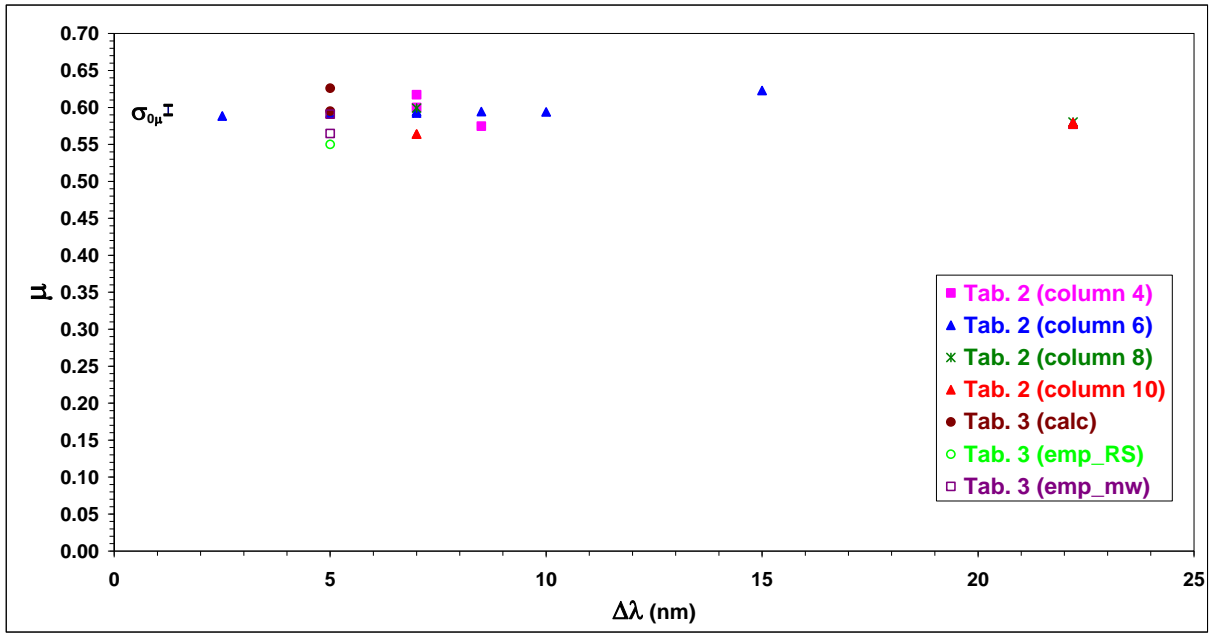

**Fig. 7**. The dependence of the empirical parameter *μ* on the half-width of the filter or on the slit width in photometric observations.

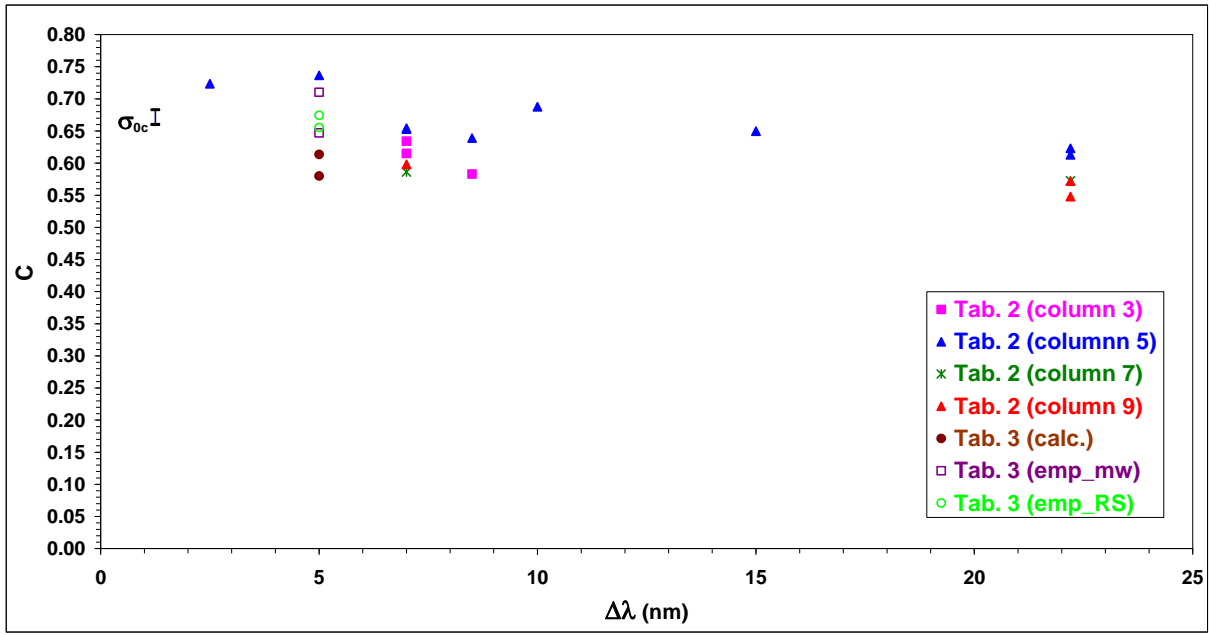

**Fig. 8**. The dependence of the empirical parameter $c$ on half-width of the filter or on the slit width in spectrophotometric observations.